\documentclass[preprint2,IOP]{emulateapj}
\usepackage{bm}
\usepackage{natbib}
\def\Pm{\mbox{\rm P}_M}
\def\Rm{\mbox{\rm R}_M}
\def\Rmc{R_{\rm crit}}

\newcommand{\be}{\begin{equation}}
\newcommand{\ee}{\end{equation}}
\newcommand{\bea}{\begin{eqnarray}}
\newcommand{\eea}{\end{eqnarray}}

\newcommand{\Eq}[1]{Eq.~(\ref{#1})}
\newcommand{\Eqs}[2]{Eqs~(\ref{#1}) and~(\ref{#2})}

\newcommand{\bra}[1]{\langle #1\rangle}
\newcommand{\bbra}[1]{\left\langle #1\right\rangle}

\newcommand{\kk}{\mbox{\boldmath $k$}}
\newcommand{\pp}{\mbox{\boldmath $p$}}
\newcommand{\aaa}{\mbox{\boldmath $a$}}
\newcommand{\AAA}{\mbox{\boldmath $A$}}
\newcommand{\q}{\mbox{\boldmath $q$}}
\newcommand{\x}{\mbox{\boldmath $x$}}
\newcommand{\y}{\mbox{\boldmath $y$}}
\newcommand{\rr}{\mbox{\boldmath $r$}}

\newcommand{\rvu}[1]{\hat{r}_{#1}}
\newcommand{\uu}{\mbox{\boldmath $u$}}
\newcommand{\RR}{\mbox{\boldmath $R$}}
\newcommand{\BB}{\mbox{\boldmath $B$}}

\newcommand{\xs}{\mbox{$\x_0$}}

\newcommand{\ys}{\mbox{$\y_0$}}

\newcommand{\rs}{\mbox{$\rr_0$}}

\shorttitle{Fluctuation dynamo at finite correlation times}
\shortauthors{Bhat and Subramanian}

\begin{document}

\title{Fluctuation dynamo at finite correlation times and the Kazantsev spectrum}

\author{Pallavi Bhat and Kandaswamy Subramanian}
\email{palvi@iucaa.ernet.in} 

\affiliation{
IUCAA, Post Bag 4, Ganeshkhind, Pune 411 007, India.
}
\date{\today}

\begin{abstract}
Fluctuation dynamos are generic to astrophysical systems. The only 
analytical model of the fluctuation dynamo
is Kazantsev model which assumes a delta-correlated in time 
velocity field. 
We derive a generalized model of fluctuation dynamo
with finite correlation time, $\tau$, using renovating flows.
For $\tau \to 0$, we recover the standard Kazantsev equation 
for the evolution of longitudinal magnetic correlation, $M_L$.
To the next order in $\tau$, the generalized equation involves
third and fourth spatial derivatives of $M_L$.
It can be recast using the Landau-Lifschitz approach, to one
with at most second derivatives of $M_L$.
Remarkably, we then find that the magnetic power spectrum,
remains the Kazantsev spectrum of $M(k) \propto k^{3/2}$,
in the large $k$ limit, independent of $\tau$.
\end{abstract}

\keywords{magnetic fields---dynamo---turbulence---magnetohydrodynamics (MHD)
---galaxies: magnetic fields}

\maketitle

\section{Introduction}

Magnetic fields are ubiquitously present in most astrophysical systems
from stars to galaxies and galaxy-clusters. They 
could be generated by dynamo amplification of weak seed fields.
A particularly generic 
dynamo is the fluctuation
or small scale dynamo \citep{Kaz68,MRS85,ZRS,KA92,KS97,RK97,
KS99,BS05,CVBLR09,federrarth11,TCB11,Suretal12,BSS12,BS13}.
Here, turbulence in a conducting plasma, 
with even a modest magnetic Reynolds number 
($\Rm > \Rmc \sim 30-500 $), 
leads to the amplification of magnetic fields on the fast eddy turn over
time scale, usually much smaller than the age of the
astrophysical system \citep{HBD04,Schek04,Scheketal05,MB10,SSBK12}. 
(Here $\Rm=u/(q\eta)$ with $u$ and $q$ respectively
characteristic velocity and wavenumber 
of the flow and $\eta$ is the resistivity.)
The $\Rmc$ depends on $\Pm = \nu/\eta$, where $\nu$ is the
viscosity and $\Rmc$ upper limit 
corresponds to $\Pm \ll 1$.
The fast growth rate implies that fluctuation dynamos 
are crucial for the early generation of magnetic fields 
in primordial stars, galaxies and galaxy-clusters. 
A clear understanding of the fluctuation dynamo
is therefore an important task. 

The only analytical treatment of the fluctuation dynamo is 
that due to \citet{Kaz68},
where the velocity field is assumed to be delta-correlated in time 
(correlation time, $\tau \to 0$). 
In this case one derives
a partial differential equation describing 
the evolution of the longitudinal magnetic correlation function, $M_L(r,t)$.
From its solutions,
Kazantsev also predicted that 
the magnetic power spectrum
for a single scale or a large $\Pm$ turbulent flow,
scales asymptotically as
$M(k) \propto k^{3/2}$, for $q \ll k \ll k_\eta$, with $k_\eta$,
the wavenumber where resistive dissipation becomes important.
This spectrum is known as the Kazantsev spectrum.
Also in the same limit, \citet{CFKV99} extended
analytic considerations to multi-point correlators, 
in a random smooth (linear) flow.

Finite-$\tau$ effects have been derived for the magnetic energy growth \citep{C97}, 
and single point PDF in the ideal limit \citep{SK01}.
\citet{KRS02} considered finite-$\tau$ correction to the two point correlator evolution, 
but seem to have kept only a subset of the terms we derive here.
\citet{MMBC11} show that 
solutions of Kazanstev equation can be made to agree with 
simulations involving finite-$\tau$ velocity flows,
if the diffusivity spectrum is approriately 
filtered out at small-scales. An analytic understanding of
the magnetic spectrum at finite-$\tau$ is however still lacking.

In this Letter, we 
give an analytic 
generalization of
the results of \citet{Kaz68} to flows
with a finite correlation time, $\tau$, by modeling the velocity 
as a renovating flow. We recover the Kazantsev evolution equation 
for $M_L$ in the $\tau\to 0$ limit, and derive 
the complete evolution equation for $M_L$ to the next 
order in $\tau$. 
We show for the first time, 
an intriguing result that the Kazantsev spectrum
is in fact preserved even for such finite-$\tau$.

\section{Fluctuation dynamo in renovating flows}

Consider the induction equation magnetic field ($\BB$) 
evolution, in a conducting fluid with velocity $\uu$,
\begin{equation}
\frac{\partial \BB}{\partial t} \;=\; \nabla \times \left( \uu \times \BB - \eta \nabla \times \BB \right).
\label{inductioneqn}
\end{equation}
We assume $\uu$ 
to have zero mean and a 
random component, which renovates every time interval $\tau$ 
\citep{DMSR84,GB92}.
It is given in the form assumed by \citet{GB92}(GB),
\begin{equation}
\uu(\x)=\aaa\sin(\q\cdot\x+\psi),
\label{uturbdef}
\end{equation}
with $\aaa\cdot\q=0$ for an incompressible flow.
In each time interval $\left[(n-1)\tau, n\tau\right]$, 
(i) $\psi$ is chosen uniformly random between 0 to $2\pi\,$; 
(ii) $\q$ is uniformly distributed 
on a sphere of radius $q=\vert\q\vert$; (iii) for every fixed $\hat{\q}=\q/q$, the direction of 
$\aaa$ is uniformly distributed 
in the plane perpendicular to $\q$. Specifically, for
computational ease, 
we modify the GB ensemble 
by choosing $a_i = P_{ij} A_j$, where 
$\AAA$ is uniformly distributed on a sphere of radius $A$, and
$P_{ij}(\hat{\q}) = \delta_{ij} - \hat{q}_i\hat{q}_j$ projects $\AAA$
to the plane perpendicular to $\q$.
Then $\bra{a^2} = 2A^2/3$. This modification in ensemble
does not affect any result using the renovating flows.
Condition (i) on $\psi$ ensures statistical homogeneity,
while (ii) and (iii) ensure statistical isotropy
of the flow.

The magnetic field evolution
in any time interval $\left[(n-1)\tau, n\tau\right]$ is
\be
B_i(\x,n\tau) \;=\; \int \mathcal G_{ij}(\x,\xs)
B_j( {\bf x_0},(n-1)\tau) \ d^3\xs
\ee
where $\mathcal G_{ij}(\x,\xs)$ is the Green's function of 
Eq.~(\ref{inductioneqn}).
To obtain $\mathcal G_{ij}(\x,\xs)$ in the
renovating flow, we use the method introduced by GB. 
The renovation time, $\tau$, is split into two 
equal sub-intervals. In the first sub-interval $\tau/2$,  
resistivity is neglected and the frozen field is advected with twice
the original velocity. In the second sub-interval, $\uu$ is neglected and the field
diffuses with twice the resistivity. 
This method, plausible in the $\tau\to 0$ limit,
has been used to recover the standard mean field dynamo equations in renovating
flows \citep{GB92,KSS12}.

In the first sub-interval $\tau/2=t_1-t_0$, 
from the advective part of Eq.~(\ref{inductioneqn}), 
we obtain the standard Cauchy solution, 
\be
B_i(\x,t_1) = \frac{\partial x_i}{\partial x_{0j}}B_j( \xs,t_0)
\equiv J_{ij}(\x(\xs)) B_j( \xs,t_0).
\ee
Here $B_j( {\bf x}_0,t_0)$ is the initial field, 
which is propagated
from $\xs$ at time $t_0$, to $\x$ at time $t_1 = t_0 +\tau/2$.  
The phase $\Phi = \q\cdot\x +\psi$ in Eq.~(\ref{uturbdef})
is constant in time as $d\Phi/dt = \q\cdot\uu =0$, from incompressibility. 
Thus $d\x/dt = 2\uu$ can be integrated to give at time $t_1 = t_0 +\tau/2$,
\be
\x = \xs +\tau \uu = \xs + \tau {\bf a} \sin(\q\cdot\xs + \psi),
\label{traj}
\ee
with the Jacobian 
\be
J_{ij}(\x(\xs)) = \delta_{ij} + \tau a_i q_j \cos(\q\cdot\xs + \psi).
\label{jacob}
\ee
It will be more convenient to work with 
the field in Fourier space, 
\be
\hat{B}_i({\bf k},t_1) = \int J_{ij}({\bf x}({\bf x_0})) B_j( {\bf x_0},t_0) e^{-i \kk\cdot\x } d^3 \x.
\label{adveceq}
\ee
In the second sub-interval ($t_1,t=t_1+\tau/2$), 
where only diffusion operates
with resistivity $2\eta$,
\be
\hat{B}_i(\kk,t) = G^{\eta}(\kk,\tau)\hat{B}_i( \kk,t_1) 
= e^{-(\eta \tau \kk^2)} \hat{B}_i( \kk,t_1),
\label{diffeq}
\ee
where $G^{\eta}$ is the 
resistive Greens function.
To derive the evolution equation for the 
magnetic two point correlation function,
we combine \Eq{adveceq} and \Eq{diffeq} to get,
\bea
&&\bra{\hat{B}_i(\kk, t)\hat{B}^*_h(\pp, t)} 
= e^{-\eta\tau(\kk^2+\pp^2)}\int \bra{J_{ij}(\xs)J_{hl}(\ys) \nonumber \\
&\times&e^{-i(\kk\cdot\x-\pp\cdot\y)}}\bra{B_j(\xs,t_0)B_l(\ys,t_0)}d^3\x d^3\y.
\label{corrlmaineq1}
\eea
Here $<.>$ denotes an ensemble average over the random velocity field
and $*$ a complex conjugate.
We have split the averaging 
between the
initial two point correlation of the magnetic field and rest of the integral,
as the initial field at $t_0$
is uncorrelated with renovating flow in the next interval $t_1-t_0=\tau/2$.

We use Eq.~(\ref{traj}) 
to transform 
from $(\x,\y)$ to $(\xs,\ys)$ in Eq.~(\ref{corrlmaineq1}). 
The Jacobian of this transformation 
is unity, due to incompressibility of the flow.
Also the initial statistical homogeneity and isotropy 
of the magnetic field are preserved at any time step.
Thus $\bra{B_j(\xs,t_0)B_l(\ys,t_0)} = M_{jl}(\vert\rs\vert,t_0)$,
where $\rs = \xs -\ys$.
Let us also
write 
$\kk\cdot\xs - \pp\cdot\ys = \kk\cdot\rs + \ys\cdot(\kk-\pp)$ 
in Eq.~(\ref{corrlmaineq1}), transform now from $(\xs,\ys)$ 
to a new set of variables $(\rs,\ys'=\ys)$, 
and integrate over $\ys'$. This leads to a delta
function in $(\kk-\pp)$ and  Eq.~(\ref{corrlmaineq1}) becomes,
\bea
&&\bra{\hat{B}_i(\kk,t)\hat{B}^*_h(\pp,t)}=(2\pi)^3 \delta^3(\kk-\pp) 
\hat{M}_{ih}(\pp,t), \nonumber \\
&&\hat{M}_{ih}(\pp, t) =  
e^{-2\eta\tau\pp^2}\int \bra{R_{ijhl}} 
M_{jl}(\rs,t_0)e^{-i\pp\cdot\rs}d^3\rs \nonumber \\ 
&&\bra{R_{ijhl}}
=\bbra{J_{ij}(\xs)J_{hl}(\ys) 
e^{-i\tau(\aaa\cdot\pp) (\sin{A} - \sin{B})}} 
\label{avterm}
\eea
where, $A=(\xs\cdot\q+\psi)$ and $B=(\ys\cdot\q+\psi)$.
We will see explicitly that $\bra{R_{ijhl}}$ is only a function
of $\rs$ as it should be from statistical homogeneity.

\section{The generalized Kazanstev equation}
 
It is difficult to evaluate $\bra{R_{ijhl}}$ exactly.
However, we motivate a 
Taylor series expansion of the exponential in  $\bra{R_{ijhl}}$ for small
Strouhl number $St = q \vert\aaa\vert \tau = q a\tau$, as follows.
Firstly $(\sin{A} - \sin{B}) 
= \sin(\q\cdot\rs/2) \cos(\psi + \q\cdot\RR_0)$,
where $\RR_0 = (\xs+\ys)/2$. Also for 
the kinematic fluctuation dynamo,
the magnetic correlation function peaks around the resistive scale 
$r_0 =\vert\rs\vert \sim 1/(q\Rm^{1/2})$,
or the spectrum peaks around $p \sim (q\Rm^{1/2})$. (Here
$p = \vert\pp\vert$ and $\Rm \sim a/(q\eta) \gg 1$). 
Thus, $q r_0 \ll 1$ and
hence $\sin(\q\cdot\rs) \sim \q\cdot\rs$. 
Subsequently the phase of the
exponential in Eq.~(\ref{avterm}) is of order 
$(p a \tau q r_0) \sim q a \tau = St$.
Thus for $St \ll 1$, one can expand the exponential
in Eq.~(\ref{avterm}) in $\tau$. We do this 
retaining terms up to $\tau^4$ order; keeping up to 
$\tau^2$ terms in Eq.~(\ref{avterm}), gives the Kazantsev equation, while the $\tau^4$ terms
give finite-$\tau$ corrections.
We get, 
\begin{eqnarray}
\bra{R_{ijhl}}&=&\bbra{H_{ijhl}[1 - i\tau\sigma \nonumber - \frac{\tau^2\sigma^2}{2!} + \frac{i\tau^3\sigma^3}{3!} + \frac{\tau^4\sigma^4}{4!}] }, \nonumber\\
&&
\label{expterms}
\end{eqnarray}
where $\sigma=(\aaa\cdot\pp)(\sin{A} - \sin{B})$ and
$H_{ijhl} = J_{ij}(\xs)J_{hl}(\ys)$ contains terms up to 
order $\tau^2$.
(We note that \citet{KRS02} 
seem to have kept only up to $p^2$ terms in \Eq{expterms}.)
To calculate $\bra{R_{ijhl}}$,
we average over $\psi$, $\hat{\bf a}$ and $\hat{\bf q}$.
Terms which are 
proportional to $\sin(...+ n\psi)$ or $\cos(...+n\psi)$ go to
zero on averaging over $\psi$. Survival of such terms which 
depend explicitly on $\xs$, $\ys$ or $\RR_0$ and
would break statistical homogeneity.
Naturally, surviving terms 
are those which depend on the relative co-ordinate $\rs$ or are 
constant.
For example $\bra{\sin A \cos A} = \bra{\sin (2\q\cdot\xs +2\psi)}/2 =0$,
while 
$\bra{\sin A \cos B} = \bra{\sin (\xs +\ys +2\psi)}/2 +
\bra{\sin(\q\cdot\rs)}/2 = \bra{\sin(\q\cdot\rs)}/2$.
Next we 
average over $\hat{\bf a}$,
by using $a_i = P_{ij}(\q) A_j$
and averaging 
independently over ${\bf A}$. The remaining $q_i$
dependent terms can be written in terms of
either $\bra{\cos(\q\cdot\rs)}$, $\bra{\cos(2\q\cdot\rs)}$ and its spatial derivatives.
Consider a simple example of the turbulent diffusion tensor,
$T_{ij} = (\tau/2) \bra{u_i(\xs)u_j(\ys)} 
=(\tau/2)\bra{a_i a_j \sin(A) \sin(B)}$, which arises 
on averaging terms proportional to $\tau^2$. 
Note that in the $\tau \to 0$ limit, $\tau$ in $T_{ij}$ 
is kept finite, 
to recover the Kazantsev equation. 
This is the reason for multiplying the velocity
two point correlator by $\tau$. We have
\bea
T_{ij} &=& 
\frac{\tau}{4}\bra{A_l A_m P_{il} P_{jm} \cos(\q\cdot\rs)}
=\frac{A^2\tau}{12}\bra{P_{ij}\cos(\q\cdot\rs)} \nonumber \\
&=&\frac{a^2\tau}{8}\left[\delta_{ij} + \frac{1}{q^2}
\frac{\partial^2}{\partial r_{0i}r_{0j}}\right]
j_0(q r_0).
\label{Tij}
\eea
Here we have used the fact that for the isotropically distributed vector
${\bf A}$, $\bra{A_i A_j} = A^2 \delta_{ij}/3$ and
the average over directions of $\q$ gives 
$\bra{\cos(\q\cdot\rs)} = j_0(q r_0)$. 

The averages of terms which are of order $\tau^4$ also introduce
the fourth order velocity correlators,
\bea
T_{mnih}^{x^2y^2}&=&\tau^2\bra{u_m(\x)u_n(\y)u_i(\x)u_h(\y)}, \nonumber \\
T_{mnih}^{x^3y}&=&\tau^2\bra{u_m(\x)u_n(\x)u_i(\x)u_h(\y)}, \nonumber \\
T_{mnih}^{x^4}&=&\tau^2\bra{u_m(\x)u_n(\x)u_i(\x)u_h(\x)}.
\label{Tijkl}
\eea
Again we multiply 
the fourth order velocity correlators
by $\tau^2$, as we envisage that 
$T_{ijkl}$ 
will be finite even in the $\tau\to 0$ limit,
behaving like products of turbulent diffusion.
Note that the renovating flow is not Gaussian random,
and hence higher order correlators of $\uu$ are not
the product of 
two-point correlators.
Interestingly, we find that the terms from \Eq{expterms} of the order 
of $\tau^3$ go to 0 on averaging.

Similarly we expand the exponential in the resistive Greens function in \Eq{avterm},
$e^{-2\eta\tau\pp^2}= 1 - 2\eta\tau\pp^2...$ and consider only 
leading order term in $\eta$, relevant 
in the independent small $\eta$ (or $\Rm \gg 1$) limit.

On combining these steps, we find that the integrand
determining the magnetic spectral tensor $\hat{M}_{ih}(\pp,t)$, 
is of the form $G(\pp)F_{ih}(\rs,t_0)$, where $G(\pp)$ is a polynomial
up to fourth order in $p_i$.
This allows for a simple inverse Fourier transform of 
$\hat{M}_{ih}(\pp, t)$, in Eq.~(\ref{avterm}) back to 
configuration space and then magnetic field correlation function
is,
\be
M_{ih}(\rr,t)
=\int G(\pp) F_{ih}(\rs,t_0)e^{i\pp\cdot(\rr-\rs)}d^3\rs 
\frac{d^3\pp}{(2\pi)^3}.
\label{Mihreal}
\ee
The various powers of $p_i$ in $G(\pp)$ above
can be written as derivatives with respect to $r_i$. 
The integral over $\pp$ then simply gives a delta
function $\delta^3(\rr-\rs)$ and this makes the integral over
$\rs$ trivial. 
Carrying out these steps the magnetic correlation function can be
written in the form 
\be
M_{ih}(\rr,t)= M_{ih}(\rr,t_0) + \tau^2 f_{ih}(\rr,t_0) 
+ \tau^4 g_{ih}(\rr,t_0)
\label{eqstruct}
\ee
We then divide 
\Eq{eqstruct} by $\tau$, take the limit of
$\tau \to 0$ and write
$(M_{ih}(\rr,t) -  M_{ih}(\rr,t_0))/\tau=\partial M_{ih}/\partial t$.
The remaining $\tau$ multiplying the term $f_{ih}$,
is absorbed into keeping $T_{ij}$ finite, 
while $\tau^2$ multiplying the term $g_{ih}$, 
is absorbed into $T_{ijkl}$, leaving 
one remaining $\tau$  as a small effective finite time parameter. 
The resulting equation for $M_{ih}$ is given by, 
\begin{widetext}
\bea
&&\frac{\partial M_{ih}(\rr,t)}{\partial t} = 2\left(-[T_{ih} M_{jl}]_{,jl} + [T_{jh} M_{il}]_{,jl} + [T_{il} M_{jh}]_{,jl} - [T_{jl} M_{ih}]_{,jl} \right) + (2T_L(0) + 2\eta) ~~\nabla^2 M_{ih} \nonumber \\
&&+ \tau \left( \left[ \tilde{T}_{mnih} M_{jl}\right]_{,mnjl} 
- 2\left[\tilde{T}_{mnrh} M_{il}\right]_{,mnrl} + \left[\left(\tilde{T}_{mnrs}
+ \frac{T_{mnrs}^{x^4}}{12}\right)M_{ih}\right]_{,mnrs} \right) 
\label{finalcor}
\eea
\end{widetext}
where $\tilde{T}_{mnih} = T_{mnih}^{x^2y^2}/{4} - T_{mnih}^{x^3y}/{3}$,
$T_L(r) = \rvu{i}\rvu{j} T_{ij}$ with $\rvu{i} = r_i/r$. 
The first line in Eq.~(\ref{finalcor}) contains exactly the terms which
give the Kazantsev equation, while the second line
contains the finite-$\tau$ corrections. 
We write
these latter terms as fourth derivative of the combined velocity and
magnetic correlators; however as both the velocity and magnetic fields
are divergence free, each spatial derivative only acts on one or the other.

Note that for a statistically homogeneous, isotropic and nonhelical
magnetic field, the correlation function 
$M_{ih} = \left(\delta_{ih} -\rvu{i}\rvu{h}\right) M_{\rm N}(r,t)
+\rvu{i}\rvu{h} M_{\rm L}(r,t)$.
Here 
$M_{L}(r,t) =\rvu{i}\rvu{h} M_{ih}$ 
and $M_{N}(r,t) = (1/2r) [\partial (r^2 M_L)/\partial r]$ are, 
respectively, the longitudinal
and transversal correlation functions of the magnetic field.
Then on contracting Eq.~(\ref{finalcor}) with $\rvu{i}\rvu{h}$
we obtain the dynamical equation for $M_L(r,t)$, the generalized Kazantsev equation,
\bea
&&\frac{\partial M_L(r,t)}{\partial t} = 
\frac{2}{r^4} \frac{\partial}{\partial r} \left( r^4 \eta_{tot} \frac{\partial M_L}{\partial r} \right) + G M_L \nonumber \\
&&+ \tau M_L^{''''} \left({\overline{T}_L} + \frac{\overline{T}_L(0)}{12}\right)
+ \tau M_L^{'''}\left(2 \overline{T}_L^{'} + \frac{8 \overline{T}_L}{r} 
+ \frac{2 \overline{T}_L(0)}{3 r} \right) \nonumber \\
&&+ \tau M_L^{''} \left(\frac{5 \overline{T}_L^{''}}{3} + \frac{11 \overline{T}_L^{'}}{r} 
+ \frac{8 \overline{T}_L}{r^2} + \frac{2 \overline{T}_L(0)}{3 r^2}\right)  \nonumber \\
&&+ \tau M_L^{'} \left(\frac{2 \overline{T}_L^{'''}}{3} + \frac{17 \overline{T}_L^{''}}{3r} 
+ \frac{5 \overline{T}_L^{'}}{r^2} - \frac{8 \overline{T}_L}{r^3} - 
\frac{2 \overline{T}_L(0)}{3 r^3} \right) 
\label{finMleq}
\eea
Here, 
$\eta_{tot} = \eta + T_L(0) - T_L(r)$ and 
$G = -2 \left(T_L^{''} + 4 T_L^{'}/r\right)$.
Also a prime denotes $\partial/\partial r$.
Furthermore, $\overline{T}_L(r) = (\overline{T}_L^{x^2y^2}/4- \overline{T}_L^{x^3y}/3)$, with 
\bea
\overline{T}_L^{x^2y^2}&=&\rvu{m}\rvu{n}\rvu{i}\rvu{h}T_{mnih}^{x^2y^2} =
-24\left(\frac{3\partial_{2z}j_0(2z)}{(2z)^3}+\frac{j_0(2z)}{(2z)^2}\right)
\nonumber \\
\overline{T}_L^{x^3y}&=&\rvu{m}\rvu{n}\rvu{i}\rvu{h}T_{mnih}^{x^3y} =
-24\left(\frac{3\partial_{z}j_0(z)}{z^3}+\frac{j_0(z)}{z^2}\right),
\label{longbarT}
\eea 
where $z=qr$ and the derivative $\partial_z$ and $\partial_{2z}$
are derivatives with respect to $z$ and $2z$ respectively.
These latter equalities give 
the explicit expressions of these 
fourth order 
correlators for the renovating flow.
Again in the limit $\tau \to 0$, 
we recover exactly
the Kazantsev equation for $M_L$.
\Eq{finMleq} allows eigen-solutions of the form
$M_L(z,t)=\tilde{M}_L(z)e^{\gamma \tilde{t}}$, 
where $\tilde{t}=t\eta_t q^2$, with $\eta_t = T_L(0)= 
a^2\tau/12 = A^2\tau/18$,
and $\gamma$ is the growth rate.
Boundary conditions are given as
$M_L'(0,t) =0$, $M_L \to 0$ as $r\to \infty$. 
Implications of the higher spatial derivative terms
are discussed below.

\section{Kazanstev spectrum at finite correlation time}

We will solve \Eq{finMleq} numerically in our detailed paper.
However, to derive both the standard Kazantsev spectrum  
in the large $k$ limit, and its finite-$\tau$ modifications, 
it suffices to go to the limit of small $z=qr\ll1$.
Expanding the Bessel functions 
in \Eqs{Tij}{longbarT} in this limit, and substituting 
$M_L(z,t)=\tilde{M}_L(z)e^{\gamma \tilde{t}}$, 
\Eq{finMleq} becomes,
\bea
&&\gamma \tilde{M}_L(z)=
\left(\frac{2\eta}{\eta_t} + \frac{z^2}{5}\right)\tilde{M}_L^{''}
+\left(\frac{8\eta}{\eta_t} + \frac{6 z^2}{5}\right) \frac{\tilde{M}_L^{'}}{z} + 2\tilde{M}_L
\nonumber \\
&&+ \frac{9 \bar\tau}{175}\left(\frac{z^4}{2}\tilde{M}_L^{''''} 
+ 8z^3 \tilde{M}_L^{'''} 
+36 z^2  \tilde{M}_L^{''}
+48 z \tilde{M}_L^{'}\right)
\label{smallrEq}
\eea
where $\bar\tau = \tau \eta_t q^2 = (St)^2/12$ and prime is now z-derivative.

Close to the origin, where $z \ll \sqrt{\eta/\eta_t}$, 
we can write $\tilde{M}_L(z) =M_0(1 - z^2/z_\eta^2)$.
Using \Eq{smallrEq}, 
$z_\eta = qr_\eta=[240/(2-\gamma)]^{1/2} [\Rm (St)]^{-1/2}$.
The $\bar\tau$ dependent terms, which are small because both
$z$ and $\bar\tau$ are small, do not affect this result.
Thus for $\Rm\gg1$, the resistive scale $r_\eta \ll 1/q$, 
although one has 
to go to sufficiently large $\Rm \gg 240/((2-\gamma)St)$ 
for this conclusion to obtain. 

Now consider the solution for $z_\eta \ll z \ll 1$.
In this limit, \Eq{smallrEq} is scale free, as 
scaling $z\to c z$ leaves it invariant. Thus
\Eq{smallrEq} has power law solutions of the form
$\tilde{M}(z) = \bar{M}_0 z^{-\lambda}$.
The appearance of higher order (third and fourth) 
spatial derivatives in \Eq{smallrEq} (or in \Eq{finMleq}), when
going to finite-$\tau$, implies that in this case,
$M_L$ evolution becomes nonlocal, 
determined by an integral type equation;
whose leading approximation for small $\bar\tau$ is 
\Eq{smallrEq}. 
However,
for small $\bar\tau$ or $St$, these higher
derivative terms only appear as perturbative terms 
multiplied by the small parameter $\bar\tau$. Then it is
possible to make 
the Landau-Lifshitz type approximation,
used in treating the effect of radiation reaction
force in electrodynamics (see \citet{LL} section 75).
In this treatment, one first ignores the perturbative terms proportional
to $\bar\tau$, which gives basically Kazantsev equation for 
$\tilde{M}_L$, and uses this 
to express $\tilde{M}_L^{'''}$ and $\tilde{M}_L^{''''}$ in terms
of the lower order derivatives $\tilde{M}_L^{''}$ and $\tilde{M}_L^{'}$.  
This gives for $z\gg z_\eta$, 
$z^3\tilde{M}_L^{'''} =  -8 z^2 \tilde{M}_L^{''} 
- z(16-5\gamma_0)\tilde{M}_L^{'}$ and 
$z^4\tilde{M}_L^{''''} = (56+5\gamma_0) 
z^2\tilde{M}_L^{''} +10(16-5\gamma_0) z\tilde{M}_L^{'}$.
Here $\gamma_0$ is the growth rate which obtains for the
Kazantsev equation in the $\tau\to 0$ limit.
Substituting these expressions back into 
the full \Eq{smallrEq} we get, 
\be
\tilde{M}_L^{''}z^2 \left( \bar{\tau}\gamma_0\frac{9}{70} + \frac{1}{5}\right) + \tilde{M}_L^{'}z \left( \bar{\tau}\gamma_0\frac{27}{35} + \frac{6}{5}\right)
+(2-\gamma)\tilde{M}_L=0
\label{smallrEq2}
\ee
Remarkably, the coefficients of the perturbative terms
in \Eq{smallrEq} 
are such that all perturbative terms which do not depend on 
$\gamma_0$ cancel out in \Eq{smallrEq2} !
Also interesting is 
the nature of the power law solution $\tilde{M}_L(z) = \bar{M}_0 z^{-\lambda}$
to \Eq{smallrEq2}.
One gets for $\lambda$,
\be
\lambda^2- 5\lambda + \frac{5(2-\gamma)}{1+\frac{9}{14}\gamma_0 \bar{\tau}} = 0;
\quad {\rm so} \ \lambda = \frac{5}{2} \pm i \lambda_I 
\label{smallrEq3}
\ee
where $\lambda_I = [20(2-\gamma)/(1 + 9\gamma_0\bar\tau/14)-25]^{1/2}/2$, 
and importantly, the real part of $\lambda$ is $\lambda_R=5/2$,
independent of the value of $\bar{\tau}$!
We can also get the approximate growth rate 
assuming $\Rm\gg1$, following
the argument from \citet{GCS96}; that one evaluates
$\gamma$ by substituting in to \Eq{smallrEq3},
the value of $\lambda=\lambda_m$ where 
$d\gamma/d\lambda =0$. This gives $\gamma_0\approx 3/4$ and
$\gamma \approx (3/4)(1 - (45/56)\bar\tau)$, which also implies 
$\lambda_I \approx 0$. (Including the effects of resistivity
gives $\lambda_I$, a small 
positive
non zero value $\propto 1/(\ln(\Rm))$
as will be shown in our detailed paper).
The $\gamma_0$ we get agrees with that of \citet{KA92},
got from looking at the evolution of $M(k,t)$.
We also note that the growth rate is reduced for a finite $\bar\tau$.
Such a reduction is found in simulations which directly compare with 
an equivalent Kazantsev model \citep{MMBC11}.

From \Eq{smallrEq3}, for 
$z_\eta \ll z \ll1$, $M_L$ is then given by
\be
M_L(z,t) = e^{\gamma\tilde{t}} \tilde{M}_0
z^{-5/2} \cos\left(\lambda_I \ln(z) + \phi\right),
\label{finsoln}
\ee
where $\tilde{M}_0$ and $\phi$ are constants.
Thus in this range, $M_L$ varies dominantly as $z^{-5/2}$,
modulated by the weakly varying cosine factor (as $\lambda_I$ is small).
Note that the magnetic
power spectrum is related to $M_L$ by
\be
M(k,t) = \int dr (kr)^3 M_L(r,t) j_1(kr) 
\label{MkMl}
\ee
The spherical Bessel function $j_1(kr)$ is peaked around
$ k\sim 1/r$, and a power law 
behaviour of 
$M_L \propto z^{-\lambda_R}$
for a range of $z_\eta \ll z=qr \ll1$,
translates into a power law
for the spectrum $M(k) \propto k^{\lambda_R -1}$ in the 
corresponding wavenumber range $q \ll k \ll q/z_\eta$. 
From the solution given in \Eq{finsoln}, we see that
in the range $z_\eta \ll z \ll1$,
$M_L$ dominantly varies as a power law with $\lambda_R=5/2$,
independent of $\tau$.
This implies remarkably that the magnetic spectrum is of the Kazantsev form 
with $M(k) \propto k^{3/2}$ 
in $k$-space, independent of $\tau$!
This is the main result of this Letter.

\section{Discussion and conclusions}

Fluctuation dynamos are
important as they ubiquitously lead to a rapid
generation of magnetic fields 
in astrophysical systems.
However their only analytical treatment, the
Kazantsev model, assumes a delta-correlated velocity field. 
Here, we have generalized 
the Kazantsev model to finite correlation 
time, $\tau$, using a velocity field which renovates every
time period $\tau$. We have shown that 
the Kazantsev equation for $M_L$ is recovered when 
$\tau \to 0$, and extended it  
to the next order in $\tau$.
In order to treat the resulting higher order (third and fourth) spatial
derivatives of $M_L$ perturbatively, we use the Landau-Lifshitz
approach; earlier used to treat the effect of the radiation reaction force.
An asymptotic treatment
shows 
firstly that the fluctuation dynamo growth rate is reduced due to 
finite $\bar\tau$. More important is
the novel and remarkable result that
the Kazantsev spectrum of $M(k) \propto k^{3/2}$, is
preserved even at finite-$\tau$. 

The finite-$\tau$ evolution equation for $M_{ih}$ (\Eq{finalcor})
or $M_L$ (\Eq{finMleq}),
is cast in terms of the general velocity 
correlators, $T_{ij}$ and $T_{ijkl}$ and matches exactly with Kazantsev equation for
the $\tau \to 0$ case. 
Morover, the forms of $T_{ij}$ and $T_{ijkl}$ at $r \ll 1/q$, are expected
to be universal due to their symmetries and divergence free properties.
These features indicate that our result on the spectrum could have a
more general validity than the context (of a renovating velocity) in which it is derived.
It would be very interesting to see if such a result 
also holds for $St \sim 1$ and to extend the finite-$\tau$ result
to helical renovating flows, issues which
we hope to address in the future. 

\acknowledgments

We thank Dmitry Sokoloff for very helpful correspondence, S. Sridhar
for several useful discussions and Axel Brandenburg and Nishant Singh
for many useful suggestions on the paper. PB thanks Nordita and 
Axel Brandenburg for support and warm hospitality while this
paper was being completed. We thank the anonymous referee for comments
which have led to improvements to our paper.

\bibliographystyle{apj}
\bibliography{reftau}

\end{document}